\documentclass[prl,aps, english, twocolumn, hyperref]{revtex4}

\usepackage{graphicx}

\usepackage{amssymb, amsmath, color, rotating}
\begin{document}

\title{Absence of damping of low energy excitations in a quasi-$2$D dipolar Bose gas}

\author{Stefan S. Natu}

\email{snatu@umd.edu}

\affiliation{Condensed Matter Theory Center and Joint Quantum Institute, Department of Physics, University of Maryland, College Park, Maryland 20742-4111 USA}

\author{S. Das Sarma}

\affiliation{Condensed Matter Theory Center and Joint Quantum Institute, Department of Physics, University of Maryland, College Park, Maryland 20742-4111 USA}

\begin{abstract}
We develop a theory of damping of low energy, collective excitations in a quasi-$2$D, homogenous, dipolar Bose gas at zero temperature, via processes whereby an excitation decays into two excitations with lower energy.  We find that owing to the nature of the low energy spectrum of a quasi-$2$D dipolar gas, such processes \textit{cannot occur} unless the momentum of the incoming quasi-particle exceeds a critical value $\textbf{k}_{\text{crit}}$. We find that as the dipolar interaction strength is increased, this critical value shifts to larger momenta. Our predictions can be directly verified in current experiments on dipolar Bose condensates using Bragg spectroscopy, and provide valuable insight into the quantum many-body physics of dipolar gases.
\end{abstract}
\maketitle

Understanding the nature of the single-particle and collective excitations of an interacting many-body system yields important insights into its macroscopic properties. For example, a hallmark of superfluidity is its ability to support dissipationless flow below a certain critical velocity, which is due to the existence of linearly dispersing excitations at low momentum \cite{landau, allum, pethick}. Following the discovery of Bose-Einstein condensation in dilute atomic vapors, several experimental groups studied the nature of the low-lying modes, and the damping of collective excitations in bosonic gases interacting with short range (contact) interactions \cite{cornellexpt1, cornellexpt2, ketterleexpt, foot, davidson}. The experiments led to important theoretical investigations on the damping of excitations in Bose condensed gases at low and high temperatures \cite{Beliaev, giorgini, liu, chung, shlyapnikov, stringari, szepfalusy, schieve, griffin}. More recently, attention has shifted to the trapping and cooling of magnetic atoms and polar molecules with \textit{long range} interactions, such as the $1/r^{3}$ dipole-dipole interaction \cite{jin}, and Bose condensates of dipolar $^{52}$Cr, $^{164}$Dy, $^{162}$Dy and $^{168}$Er have already been created in the laboratory \cite{pfau, pfau2, dysprosium, dysprosium2, erbium}. Motivated by these developments, we describe a theory of damping of low energy excitations in a quasi-$2$D dipolar Bose condensate, finding an intriguing damping process which switches on only above a certain critical wave-number, implying that the low lying collective excitations in this system are undamped. 

The physics of dipolar Bose gases in the continuum is qualitatively different from that of a Bose gas with short range interactions. This is largely due to the anisotropic nature of the long range, dipole-dipole interaction, which introduces novel phenomena such as geometry dependent mechanical stability \cite{pfau2}, d-wave collapse \cite{pfau}, anisotropic critical velocity for dissipationless flow \cite{wilson} and a roton-maxon dispersion relation in quasi-$2$D systems \cite{santos, lahayereview}, analogous to He--$4$ \cite{marisreview, mills, griffinbook}.  Here we theoretically predict yet another novel property of a quasi-$2$D dipolar Bose gas at zero temperature: \textit{the absence of damping for long wavelength collective excitations}. We show that the decay of a single excitation into two excitations with lower energy (Beliaev damping) is energetically forbidden in a quasi-$2$D dipolar gas. Furthermore, this effect is present even for weak dipolar interactions, and can thus be observed in current experiments, without the use of Feshbach resonances. 

The Hamiltonian for a uniform dipolar Bose gas takes the form:
\begin{eqnarray}\label{Ham}
{\cal{H}} = \int d\textbf{r}~\Psi^{\dagger}(\textbf{r}, t)\Big[-\frac{\hbar^{2}\nabla^{2}}{2m} - \mu \Big]\Psi(\textbf{r}, t) + \\\nonumber \frac{1}{2}\int d\textbf{r}d\textbf{r}^{'}V_{\text{tot}}(\textbf{r}-\textbf{r}^{'})\Psi^{\dagger}(\textbf{r}, t)\Psi^{\dagger}(\textbf{r}^{'}, t)\Psi(\textbf{r}^{'}, t)\Psi(\textbf{r}, t)
\end{eqnarray}
where $\Psi(\textbf{r}, t)$ is the bosonic annihilation operator at position $\textbf{r}$ and time $t$, and $\mu$ is the chemical potential. The interaction potential is assumed to be a sum of two terms $V_{\text{tot}}(\textbf{r} -\textbf{r}^{'}) = g~\delta(\textbf{r}-\textbf{r}^{'}) + V_{\text{dip}}(\textbf{r}-\textbf{r}^{'})$, where $g = 4\pi\hbar^{2}a/m$ parametrizes the contact part of the potential, where $a$ is the s-wave scattering length, and $V_{\text{dip}}(\textbf{r}-\textbf{r}^{'})$ is the dipolar potential. For a gas with dipole moment $d$, where all the dipoles are oriented along the $z$ axis, $V_{\text{dip}}(\textbf{r}-\textbf{r}^{'}) = d^{2}(1 - 3\cos^{2}(\theta))/|\textbf{r} -\textbf{r}^{'}|^{3}$, where $\theta$ is the angle between the vector $\textbf{r} -\textbf{r}^{'}$ and the $z$ axis.

For atomic dipolar gases such as $^{52}$Cr \cite{pfau}, $^{164}$Dy \cite{dysprosium2}, $^{162}$Dy \cite{dysprosium} and $^{168}$Er atoms \cite{erbium}, that have already been Bose condensed in the laboratory, the dipolar and contact parts of the interaction are comparable to one another $0.1 < g_{d}/g \lesssim 1$. Stronger dipole-dipole interactions, $g_{d}/g \sim 10$ can be achieved using polar molecules (or by using a Feshbach resonance to tune $g$ close to zero \cite{pfau}), which are currently being trapped and cooled by several groups \cite{jin}.

Here we work in a quasi-two dimensional geometry, which can be implemented in harmonically trapped gases with a trapping potential of the form $U(\textbf{r})  = \frac{1}{2}m(\omega^{2}_{x}x^2 + \omega^{2}_{y}y^2+\omega^{2}_{z}z^2)$, where $\omega_{z} \gg \omega_{x}, \omega_{y}$, or by using a deep optical lattice in the $z$-direction. In the limit $\mu \ll \hbar\omega_{z}$, with no loss of generality, the density can be expressed as $n(\textbf{r} = \{\rho, z\}) = |\Psi(\rho, z)|^{2} =n^{\text{2D}}(\rho)\Psi(z) = \frac{1}{\sqrt{\pi l^{2}_{z}}}n^{\text{2D}}(\rho)e^{-z^{2}/l^{2}_{z}}$, where $\rho$ and $z$ is the radial and the axial co-ordinate respectively, and $l_{z}$ is a length-scale on the order of the harmonic oscillator wave-length in the axial direction $l_{z} \sim \sqrt{\hbar/m \omega_{z}}$ \cite{fischerdipole}. Integrating out the $z$-direction, one obtains an effective two dimensional description, which depends on $\rho$ alone. For the homogeneous case we consider ($\omega_{x} = \omega_{y} = 0$), the Fourier transform of the resulting quasi-$2$D interaction potential reads \cite{fischerdipole, wilson}:
\begin{equation}\label{q2dpot}
V^{\text{q2D}}(k) = \frac{1}{\sqrt{2\pi}l_{z}}\Big(g + g_{d}F\Big(\frac{{k}l_{z}}{\sqrt{2}}\Big)\Big)
\end{equation}
where $k = \sqrt{k^{2}_{x}+k^{2}_{y}}$ is the radial momentum, $g_{d} = \frac{8\pi}{3}d^{2}$, and $F(x) = 1 - \frac{3}{2}\sqrt{\pi}x~\text{Erfc}(x)e^{x^{2}}$, where $\text{Erfc}(x)$ is the complimentary error function. When the dipoles are aligned parallel to one another,  the dipole-dipole interaction only depends on the magnitude of the radial momentum. 

Strictly speaking, the Gaussian model used above is only valid for $\mu \ll \hbar\omega_{z}$. Santos \textit{et al.} \cite{santos} use a more rigorous approach, they model the axial extent of the cloud using a Thomas-Fermi distribution, and numerically integrate over the $z$ direction to obtain a more accurate quasi-$2$D potential. We use Eq.~\ref{q2dpot} here as it correctly captures all the qualitative features of the true quasi-$2$D dipolar potential, and allows us to obtain some analytic results, thus serving as a conceptual guide for theory and experiment \cite{lahayereview}. 
This advantage of the Gaussian approximation compensates for the slight quantitative inaccuracy incurred in our treatment, and if future experiments demand quantitative refinement, a numerical analysis using better wave-functions should be straightforward to carry out within our approach.

At temperatures below the Bose condensation temperature, we may write $\Psi$ as a sum of condensate and non-condensate wave-functions, 
$\Psi(\rho, t) = \phi(\rho, t) + \psi(\rho, t)$, where $\phi(\rho, t) = \langle \Psi(\rho, t) \rangle$ represents the condensate field, and $\psi(\rho, t)$ represents the non-condensed atoms, which by definition have the property $\langle \psi(\rho, t) \rangle = 0$ \cite{giorgini}. The condensate field obeys the usual Gross-Pitaevskii equation which can be solved to yield the condensate density in equilibrium $n^{\text{2D}}_{0} = \phi^{2}_{0}$. In a homogeneous gas, $\phi_{0}$ is independent of $\rho$. 

The excitation spectrum is found by writing $\Psi(\rho, t) = \phi_{0} +\psi(\rho, t)$ in Eq.~\ref{Ham},  and ignoring terms proportional to $\psi^{3}$ and $\psi^{4}$, to obtain a quadratic Hamiltonian in terms of the non-condensed fields. The resulting Hamiltonian can be diagonalized in momentum space via the usual Bogoliubov transformation: $\psi(\rho, t) = \sum_{\textbf{k}}u_{\textbf{k}}(\rho)a_{\textbf{k}}(t) +v^{*}_{\textbf{k}}(\rho)a^{\dagger}_{\textbf{k}}(t)$ \cite{pethick}, where $\textbf{k} = \{k_{x}, k_{y}\}$, $a_{\textbf{k}}(t)$ denotes the bosonic annihilation operator for a quasi-particle with momentum $\textbf{k}$ at time $t$. The complex numbers $u_{\textbf{k}}(\rho) = u_{k}e^{i\textbf{k}.\rho}$ and $v_{\textbf{k}}= v_{k}e^{i\textbf{k}.\rho}$ obey $|u_{k}|^{2} - |v_{k}|^{2} =1$. The Bogoliubov Hamiltonian reads  \cite{fischerdipole, goral}: ${\cal{H}}_{0} = \sum_{\textbf{k}}E_{k}a^{\dagger}_{\textbf{k}}a_{\textbf{k}}$,
where $E_{k} = \sqrt{\epsilon_{k}(\epsilon_{k} + 2 V^{\text{q2D}}(k)n^{\text{2D}}_{0})}$ is the quasi-particle energy obtained by solving the Bogoliubov equations with the constraint on $u_{k}$ and $v_{k}$. Here $\epsilon_{k} = \hbar^{2}k^{2}/2m$ is the single-particle energy, and $u_{k} = \sqrt{\frac{1}{2}\Big(\frac{\epsilon_{k} + V^{\text{q2D}}(k)n^{\text{2D}}_{0}}{E_{k}} + 1\Big)}$ and $v_{k} = -\text{sgn}(V^{\text{q2D}}(k))\sqrt{\frac{1}{2}\Big(\frac{\epsilon_{k} + V^{\text{q2D}}(k)n^{\text{2D}}_{0}}{E_{k}} - 1\Big)}$ are the Bogoliubov coefficients.

The $3$D dipolar gas is susceptible to a collapse instability for $g < g_{d}$, (signaled by the appearance of imaginary frequencies $E_{k}$ as $k \rightarrow 0$) due to the partially attractive nature of the dipole-dipole interaction \cite{goral}. This was demonstrated experimentally in dipolar $^{52}$Cr atoms by Lahaye \textit{et al.} \cite{pfau}, where a Feshbach resonance was used to tune the s-wave scattering length near the zero crossing. However, as was pointed out by Fischer \cite{fischerdipole}, the quasi-$2$D dipolar potential given by $V^{\text{q2D}}$ is energetically stable, even in the absence of a repulsive contact interaction.

Below we illustrate how the damping of excitations in a purely dipolar gas ($g = 0$ in Eq.~\ref{q2dpot}) differs from that of a gas with purely contact interactions, which has been well understood for some time \cite{giorgini, chung, liu, Beliaev}. We then compute the damping rate in a gas with the full interatomic potential of Eq.~\ref{q2dpot} to make predictions relevant to current experiments.

To leading order, the damping of excitations arises due to coupling between the fluctuations of the condensate and the non-condensed fields. At zero temperature, excitations decay via Beliaev damping, where a particle in the condensate and a quasi-particle with momentum $\textbf{p}$ are annihilated (created), and two quasi-particles with momenta $\textbf{k}$ and $\textbf{q}$ are created (annihilated) \cite{Beliaev}. 

Generally speaking for a $d-$dimensional gas ($d \geq 2$) interacting with a potential $V_{\text{tot}}(\textbf{r} - \textbf{r}^{'})$, such damping processes result from an interaction Hamiltonian \cite{Beliaev}:
\begin{equation}\label{hint} 
{\cal{H}}_{int} \sim \int d\textbf{r}~d\textbf{r}^{'} V_{\text{tot}}(\textbf{r} - \textbf{r}^{'})(\phi^{*}_{0}(\textbf{r})~\psi^{\dagger}(\textbf{r}^{'})\psi(\textbf{r}^{'})\psi(\textbf{r}) + \text{h.c})
\end{equation}

Treating ${\cal{H}}_{int}$ to first order in perturbation theory for a homogeneous gas at zero temperature, and following the approach of Giorgini \cite{giorgini}, we find that the Beliaev damping rate takes the form:
\begin{equation}\label{bdamping}
\Gamma_{\text{B}}(\textbf{p}) = \frac{\pi n_{0}}{2\hbar}\sum_{\textbf{k},\textbf{q}}\delta(\epsilon_{\textbf{p}} - (\epsilon_{\textbf{k}} +\epsilon_{\textbf{q}}))  \delta_{\textbf{p}, \textbf{k}+\textbf{q}}A_{\textbf{k}, \textbf{q}}^{2}
\end{equation} 
where $\textbf{p}$ is the momentum of the annihilated (created) quasi-particle and $\textbf{k}$ and $\textbf{q}$ are the momenta of the created (annihilated) quasi-particles in $d$-dimensions. Here $n_{0}$ is the condensate density. The $\delta_{\textbf{p}, \textbf{k}+\textbf{q}}$ enforces conservation of momentum, and the matrix element $ A_{\textbf{k}, \textbf{q}}$ is given by:
\begin{widetext}
\begin{eqnarray}\label{matelem}
A_{\textbf{k}, \textbf{q}} =u_{k+q}\Big\{(V_{\text{tot}}(\textbf{q}) +  V_{\text{tot}}(\textbf{k}))u_{k}u_{q} + V_{\text{tot}}(\textbf{q}+\textbf{k})(u_{k}v_{q} + v_{k}u_{q}) +  u_{k}v_{q} V_{\text{tot}}(\textbf{q}) + V_{\text{tot}}(\textbf{k})u_{q}v_{k}\Big\} + \\\nonumber v_{k+q}\Big\{(V_{\text{tot}}(\textbf{q}) + V_{\text{tot}}(\textbf{k}))v_{k}v_{q} + V_{\text{tot}}(\textbf{q}+\textbf{k})(u_{k}v_{q} + u_{q}v_{k}) + V_{\text{tot}}(\textbf{q})u_{k}v_{q}  + V_{\text{tot}}(\textbf{k})u_{q}v_{k}\Big\}
\end{eqnarray}
\end{widetext}
 which is manifestly symmetric upon interchange of $\textbf{q}$ and $\textbf{k}$. 
 
We emphasize that in deriving Eqs.~\ref{bdamping} and \ref{matelem}, we simply assume that  $V_{\text{tot}}(\textbf{r}) = V_{\text{tot}}(-\textbf{r})$, and as such these equations describe Beliaev damping in a gas with \textit{arbitrary} long range interactions, not just dipolar interactions. A detailed derivation of this result and its extension to finite temperature will be published elsewhere \cite{naturyaninprep}. 

For a $3$D gas interacting with purely contact interactions ($V_{\text{tot}}(\textbf{k}) = g$), one readily checks that Eq.~\ref{matelem} reduces to the familiar expression for the Beliaev damping rate obtained by Giorgini \cite{giorgini}. For a quasi-$2$D dipolar gas, we replace $n_{0}$ in Eq.~\ref{bdamping} with $n^{\text{2D}}_{0}$ and $V_{\text{tot}}$ in Eq.~\ref{matelem} with $V^{\text{q2D}}$ given by Eq.~\ref{q2dpot}, and the sums are performed over two-dimensional momenta.

From Eq.~\ref{bdamping}, it is clear that Beliaev damping is only possible if the conditions for energy and momentum conservation can be simultaneously satisfied. For a Bose gas with contact interactions ($g_{d} = 0$), the low energy dispersion takes the form $E_{p} \approx c p + \frac{p^{3}}{8m^{2}c}$, where $c = \sqrt{gn^{\text{2D}}_{0}/\sqrt{2\pi}m l_{z}}$ is the phonon velocity. At low energies this dispersion is analogous to that of He--$4$ \cite{marisreview}. As was pointed out by Maris \cite{marisreview}, a quasi-particle with momentum $\textbf{p}$ can decay into two phonons of lower energy provided the energy $E_{p}$ lies above the phonon dispersion curve ($E_{p} = cp$). For a gas with contact interactions, the low energy dispersion bends upwards (due to the positive co-efficient of the $p^{3}$ term), and Beliaev damping is allowed by energy-momentum conservation at arbitrary wave-vectors \cite{giorgini, liu, chung}. 

 By contrast, the low energy dispersion for a quasi-$2$D, purely dipolar Bose gas ($g = 0$) reads:
\begin{equation}\label{edisplow}
E_{p} \approx c_{d} p - \frac{3\sqrt{\pi}l_{z}}{4\sqrt{2}\hbar}c_{d}p^{2}  + \frac{p^{3}}{8m^{2}c_{d}}\Big(1 + (6-\frac{9\pi}{8})\frac{m^{2}c^{2}_{d}l^{2}_{z}}{\hbar^{2}}\Big)
\end{equation}
where $c_{d} = \sqrt{\frac{g_{d}n^{\text{2D}}_{0}}{\sqrt{2\pi} m l_{z}}}$ is the phonon velocity for the dipolar gas. In the weakly interacting gas $g_{d}n^{\text{2D}}_{0}/l_{z} \ll \hbar^{2}/ml^{2}_{z} = \hbar\omega_{z}$, the second term in the brackets proportional to $p^{3}$ can be ignored. 

As Eq.~\ref{edisplow} shows, the quadratic term in the expansion of the energy \textit{always} dominates the cubic term for momenta $p < p_{\text{crit}}(c_{d}) = 3\sqrt{2\pi}m^{2}l_{z}c^{2}_{d}/\hbar$. For momenta smaller than $p_{\text{crit}}$, a quasi-particle with momentum $\textbf{p}$ will always have lower energy than two phonons with momenta $\textbf{q}$ and $\textbf{p}-\textbf{q}$ for all $0 < q < p$. Beliaev damping is thus forbidden by energy and momentum conservation. For momenta larger than $p_{\text{crit}}$, the cubic term dominates the quadratic term in the low energy expansion, and Beliaev damping is allowed. 

\begin{figure}
\begin{picture}(200, 100)
\put(0, -10){\includegraphics[scale=0.45]{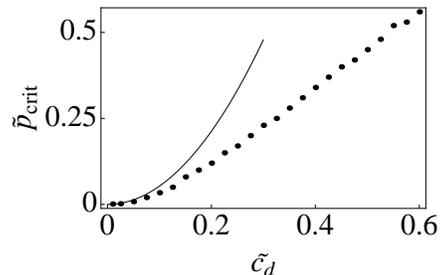}}
\end{picture}
\caption{\label{critmom}The data points are the numerically obtained values of the dimensionless $\tilde p_{\text{crit}} = p_{\text{crit}}/(\hbar\sqrt{2}/l_{z})$ plotted as a function of the dimensionless  sound velocity $\tilde c_{d} = c_{d}/\omega_{z}l_{z} = \sqrt{g_{d}n^{\text{2D}}_{0}/\sqrt{2\pi} l_{z}\hbar\omega_{z}}$, defined as the ratio of the interaction energy ($E_{\text{int}} = g_{d}n^{\text{2D}}_{0}/\sqrt{2\pi}l_{z}$) to the harmonic oscillator energy in the $z-$direction ($E_{\text{ho}} = \hbar \omega_{z}$): $\tilde c_{d} = \sqrt{E_{\text{int}}/E_{\text{kin}}}$. Collective modes with momenta below $p_{\text{crit}}$ are undamped. The solid line is the analytic result: $\tilde p_{\text{crit}} = 3\sqrt{\pi}\tilde c_{d}^{2}$ obtained using Eq.~\ref{edisplow} which works well at small $\tilde c_{d}$. For larger $\tilde c_{d}$, Eq.~\ref{edisplow} underestimates the dipolar dispersion leading to a larger $p_{\text{crit}}$. The contact part of the interaction is set to zero.}
\end{figure}

In Fig.~\ref{critmom} we plot $p_{\text{crit}}$ as a function of $c_{d}$ obtained by numerically solving for the energy conservation constraint. Note that the momenta are expressed in units of $\hbar\sqrt{2}/l_{z}$ and a dimensionless speed of sound is defined as the ratio of the mean-field interaction energy $E_{\text{int}} = g_{d}n^{\text{2D}}_{0}/\sqrt{2\pi}l_{z}$ to the harmonic oscillator energy in the confining direction $E_{\text{ho}} = \hbar\omega_{z}$: $\tilde c_{d} = \sqrt{E_{\text{int}}/E_{\text{ho}}}$.
The solid line shows the analytic result, which is valid for small $c_{d}$. For stronger dipole-dipole interactions, Eq.~\ref{edisplow} underestimates the dispersion leading to a larger $p_{\text{crit}}$.  

We emphasize that the physics here is different from that of He--$4$ where phonon damping can always occur via the Beliaev mechanism \cite{mills, marisreview, griffinbook}. In this system, the low energy dispersion can be parameterized as $E(p) \sim \alpha p + \beta p^{3} - \gamma p^{5}$ (where $\alpha$, $\beta$ and $\gamma > 0$  \cite{marisreview}). It is only at intermediate values of momentum, where the $\gamma p^{5}$ term becomes relevant, that excitations are stable against Beliaev decay. Similar physics also occurs in optical lattices where the band structure modifies the dispersion close to the Brillouin zone boundary \cite{griffin1, griffin2}.

By contrast, for the quasi-$2$D dipolar gas, the $p^{2}$ term in the dispersion dominates the $p^{3}$ term for \textit{arbitrarily} weak $p$. This leads to a complete turning off of Beliaev damping at low momenta. We are not aware of any other bosonic system with this feature. For larger values of the dipole-dipole interaction strength, the dispersion develops a roton-maxon feature \cite{santos}, and a quasi-particle (maxon) can decay into two rotons \cite{pitaevskii, naturyaninprep2}.  

\begin{figure}
\begin{picture}(200, 90)
\put(10, -10){\includegraphics[scale=0.4]{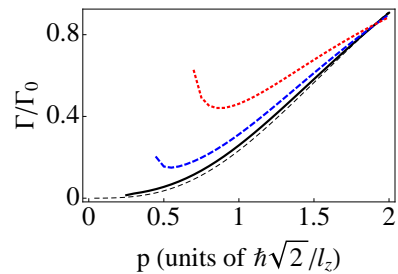}}
\end{picture}
\caption{\label{dampingrate} (Color Online) Damping rate (normalized to $\Gamma_{0} = (g/\sqrt{2\pi }l_{z})^{2}n^{\text{2D}}_{0}m/4\pi \hbar^{3}$) in a  Bose gas with dipolar and contact interactions, plotted versus $p$. The thin dashed curve is the result for purely contact interactions ($\tilde g = 0$) (c.f Refs.~\cite{chung, giorgini}) shown for comparison. The damping rate scales as $p^{3}$ at low momenta. The remaining curves show the damping rates in a gas with dipolar interactions: $\tilde g = g_{d}/g = 0.1$ (solid, black), $\tilde g= 0.25 $ (blue dashed curve starting from $p = 0.45 \hbar\sqrt{2}/l_{z}$) and $\tilde g = 0.5 $ (red dotted curve starting from $p = 0.7 \hbar\sqrt{2}/l_{z}$). For a dipolar gas, there is no damping until $p$ reaches a threshold value (See Fig.~\ref{critmom}). The dip in the damping rate at intermediate momenta is due to the appearance of a slight shoulder in the dispersion relation at intermediate $\tilde g$ \cite{santos}.} 
\end{figure}

In Fig.~\ref{dampingrate}, we plot the calculated Beliaev damping rate as a function of $p$ for a gas with dipolar plus contact interactions (Eq.~\ref{q2dpot}) obtained by integrating Eq.~\ref{bdamping} using Eq.~\ref{matelem}. The contact part of the interaction is held fixed  (we choose $\tilde c = c/l_{z}\omega_{z} = 1$) and the dimensionless ratio of the dipole-dipole interaction strength to the contact interaction $\tilde g = g_{d}/g$ is varied. The damping rate is normalized by $\Gamma_{0} = (g/\sqrt{2\pi }l_{z})^{2}n^{\text{2D}}_{0}m/4\pi \hbar^{3}$.  

For a gas with purely contact interactions ($\tilde g = 0$), shown as the thin, solid line, it is well known that for small momenta $p \ll 2mc$, the Beliaev damping rate scales as $p^{2d-1}$ where $d$ is the dimension \cite{Beliaev, liu, giorgini, chung}. To see this, note that to leading order $A_{\textbf{k}\textbf{q}}$ scales as \cite{giorgini, liu, chung}: $A^{2}_{\textbf{k}\textbf{q}} \sim \frac{p |\textbf{p}-\textbf{q}| q}{c^{3}}$. Inserting this expression into Eq.~\ref{bdamping} and using the condition for small angle scattering: $|\textbf{p} - \textbf{q}| \approx p-q + \frac{pq\theta^{2}}{2(p-q)}$ where $\theta = \frac{\sqrt{3}(p-q)}{2mc}$ \cite{giorgini, liu, chung}, one obtains the rate of Beliaev damping at small momenta. Similar scaling also occurs for damping of low energy magnons in anti-ferromagnets \cite{magnon}. 


As Fig.~\ref{dampingrate} shows, the behavior is strikingly different for a gas with dipolar interactions. For a dipolar gas, there is no Beliaev damping at low $p$, so the damping rate is zero until $p$ reaches a threshold momentum. Beyond this the damping rate jumps  to a finite value. The value of this threshold momentum  and the corresponding jump in the damping rate increases with increasing $\tilde g$. At large $p$, the contact plus dipolar dispersion becomes free particle-like and the behavior is similar to that of a gas with purely contact interactions.

For weak dipolar interactions (solid black line in Fig.~\ref{dampingrate}, the damping rate closely follows that of the gas with purely contact interactions. For larger values of $\tilde g$ (shown as the blue dotted curve and the red dashed curve), we find a dip in the damping rate at intermediate momenta, which moves to larger $p$ as $\tilde g$ is increased. The size of the dip increases with increasing $\tilde g$. The location of the dip is consistent with the location of the shoulder in the dispersion relation at intermediate $p$ and intermediate $\tilde g \sim 0.5$. This shoulder is the precursor of the roton minimum which becomes more pronounced at large values of $\tilde g$ \cite{santos}. (For $\tilde c = 1$, the roton minimum develops at $\tilde g \approx 5$). It is well known that rotons are stable against Beliaev decay \cite{pitaevskii, marisreview}, so we expect the damping rate to completely vanish for momenta close to the roton minimum, as $\tilde g$ is increased further. A quantitative study of the damping rate for the maxon-roton excitations will be the subject of a future work \cite{naturyaninprep2}.

We now briefly discuss the possibility of observing the phenomenon discussed here in experiments. A key challenge is that the discretization of the low energy spectrum in trapped gases renders Beliaev damping inactive \cite{shlyapnikov}. Nonetheless, Hodby \textit{et al.} \cite{foot} observed Beliaev damping by carefully designing the trap geometry so as to transfer energy between two low lying collective modes. Katz \textit{et al.} \cite{davidson}  used Bragg spectroscopy to probe Beliaev damping in a $3$D Bose gas with short-range interactions. In liquid He-$4$, the collective spectrum was studied using neutron scattering, which measures the dynamic structure factor $S(\textbf{q}, \omega) = \int dt e^{-i\omega t}\langle \rho_{\textbf{q}}(t) \rho_{-\textbf{q}}(0)\rangle$ (where $\rho_{\textbf{q}}(t) = \sum_{\textbf{k}}a^{\dagger}_{\textbf{k}}(t)a_{\textbf{k}+\textbf{q}}(t)$) (See Ref.~\cite{cowley} and references therein). The damping rate can be extracted from the data using sum rules \cite{griffinbook}. Recently developed high resolution imaging methods can directly probe the time evolution of the \textit{static} structure factor $S(\textbf{q}, t) = \langle \rho_{\textbf{q}}(t) \rho_{-\textbf{q}}(t)\rangle$ following a sudden quench in the interaction \cite{chengstruc, chengsakh}. However, quantitatively relating the features in the data to the damping of quasi-particles is still an open problem.

In conclusion, we develop a theory for damping of collective excitations in dipolar gases at zero temperature, generalizing existing works on damping in gases with contact interactions. Focusing on the long wavelength, low energy limit, we find that the nature of the dispersion forbids the decay of a quasi-particle into two quasi-particles with lower energy (Beliaev damping). A direct experimental verification of our predictions should be possible with existing dipolar systems. 

\textit{Acknowledgements.---} It is a pleasure to thank Ryan M. Wilson for numerous discussions and for his careful reading of the manuscript. This work is supported by JQI-NSF-PFC, AFOSR-MURI, and ARO-MURI.

\end{document}